# Predicting Formula 1 Race Outcomes: Decomposing the Roles of Drivers and Constructors through Linear Modeling

Saurabh Rane (rane.saurabh7@gmail.com, saurabhr.com)

July 31st, 2025

# Abstract

Formula 1 performance is a combination of the car's ability and the driver's ability. While a given race or season can tell you how well a car and driver performed jointly, isolating the individual impact of the driver and constructor remains challenging.

This paper extends a Regularized Adjusted Plus Minus (RAPM) methodology (Sill 2010), commonly used in basketball and hockey, to parse out individual driver and constructor impact. It employs a time-decayed ridge regression with LOESS (Jacoby 2000) smoothing to predict race results for the Hybrid Engine Era (2014 – 2024). By measuring the constructor and driver coefficients over time, we measure the relative individual impact of driver and constructor throughout the period.

Results show that constructors explain 64.0% of the variance in race outcomes in the Hybrid Engine Era. Additionally, constructors have increased importance in benchmarked rank-agnostic cohorts (e.g., Top 10 points finishers) and decreased importance in qualifying.

By decomposing performance into individual driver and constructor metrics, we create a robust framework for inter-constructor driver comparisons that the Formula 1 points system obfuscates. Our work enhances the understanding of driver and constructor contributions to race success, offering valuable insights for strategic decision-making in Formula 1.

# 1. Introduction

The realm of sports often celebrates the exceptional athletic prowess of individuals or teams, and Formula 1 is no exception. Victorious drivers are traditionally lauded with champagne showers and podium ceremonies. However, the success of a driver is not solely attributable to their skills. As 2016 World Drivers' Champion Nico Rosberg remarked, success in Formula 1 is "20 percent driver and 80 percent car" (Spurgeon, 2009).

The significance of the car's role is evident in the historical data. Since 2000, the World Constructors' Champion has only differed thrice from the team of the World Drivers' Champion (Formula 1). Moreover, in the eleven completed seasons of the Hybrid era, the World Drivers' Champion and runner-up have hailed from the same team on six occasions.

While it is well-acknowledged that constructors play a pivotal role in Formula 1 success, there is a need to delve deeper into the individual variances attributable to constructors and drivers. Prior analyses (Kesteren, 2023; Rockerbie, 2021; Bell, 2016) have explored the overall variance explained by constructors versus drivers, assigning "skill" values to each.

Our objective is to quantify the individual impacts of constructors and divers via a predictive model that forecasts future races accurately. Current comparisons of teams (inter-constructor) and drivers within the same team (intra-constructor) are straightforward through Formula 1's point system. However, comparing drivers across different teams (inter-constructors) remains challenging due to the dominant influence of the constructor.

Moreover, our approach differs from the existing literature. Bell (2016) and Kesteren (2023) have produced the most similar work in the field. Bell's methodology treats a driver's entire career as a single entity, while Kesteren's analysis is conducted at the season level. In contrast, our model evaluates driver and constructor ratings at a race-level granularity. This finer level of detail allows for a more responsive assessment of driver and constructor performance, but increased sensitivity introduces greater variance and results in a more volatile trajectory of ratings.

Finally, we seek to enhance our understanding of the relative contributions of constructors and drivers in determining race outcomes over time, providing a more nuanced analysis of their impacts on race rankings.

## 2. Data Processing

We collect data from all Formula 1 races through the Ergast API and Jolpica-F1project via the f1dataR R package (Ergast, f1dataR). Our analysis primarily focuses on the hybrid engine era (2014 to 2024), utilizing race results from 2012 onwards to provide a warm start.

For our primary model, we exclude non-finishers (DNF) to avoid the additional noise and complexity they introduce to the outcome of interest, as similarly noted by Kesteren (2023).

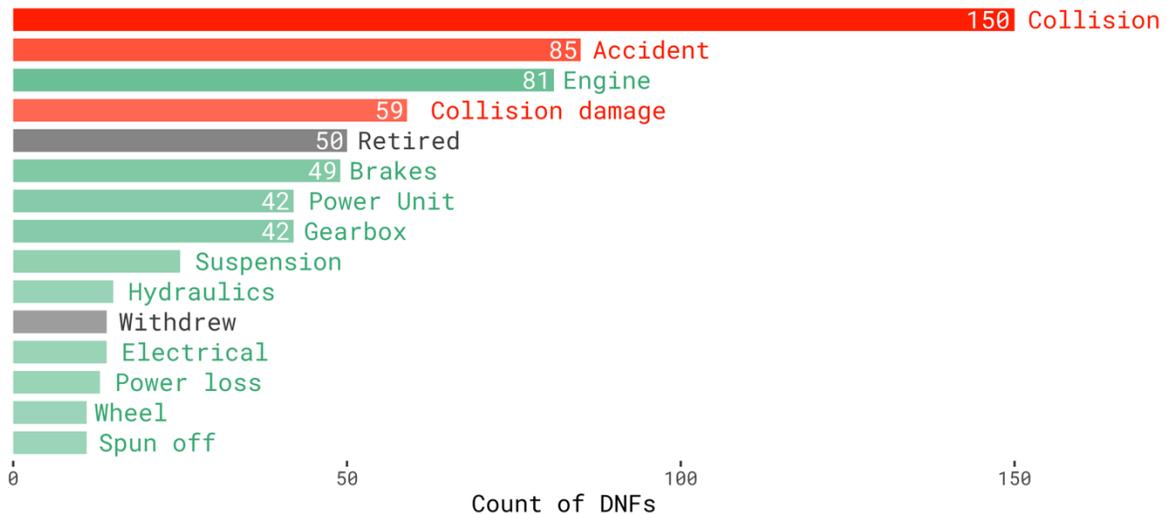

*Figure 1 Top DNF reasons from 2014 to 2024. DNFs are categorized as driver-related, constructor-related, or indeterminate based on the author's discretion.*

As seen in Figure 1, there's a myriad of DNF reasons, with the plurality coming from accident or collisions. However, assigning the root cause of certain DNFs, particularly driver-driver collisions, can be challenging. For example, in the 2024 Canada GP, both Albon and Sainz had their DNFs classified as *Collision*. This would result in penalizing both drivers and constructors in the proposed rating system despite the collision being driven by Sainz and Ferrari. Additionally given the time-decay and point-in-time nature of our study, indexing on DNFs may disproportionately add negative noise to a driver or constructor coefficient.

For completeness, we also test model variants that include DNFs wholly and another that partially assigns DNF cause to either driver or constructor based on the type of DNF. DNFs categorized as Collision, Accident, or Spun-Off are attributed to driver fault, while other causes are attributed to constructor fault.

The features used in the model are binary indicators for the parent constructor and driver at the race level. The term "parent constructor" refers to constructors that have undergone name changes but remain the same underlying entity. For example, Toro Rosso, Alpha Tauri, and RB are considered the same constructor in this analysis.

The response variable is the rank finish of the driver-constructor pair in each race. Points-scoring positions (ranks 1 to 10) are weighted equally, as drivers are expected to compete equally for these positions. Positions beyond 10 are weighted on a linear decay until the last position, reflecting the expectation that the effort expended for lower places is probably less than that for the top places (Henderson, 2021). The calculation of rank finishing weights is seen below in (1).

$$weight_{Rank\ Finish} = \begin{cases} 1, & Position \leq 10 \\ \dfrac{Cars_{Outside\ Points} - (Position - 9)}{Cars_{Outside\ Points} + 1}, & Position > 10 \end{cases}$$

$$Cars_{Outside\ Points} = Cars_{Total} - 10$$

(1)

## 3. Methods

### 3.1. Linear Model with Ridge Regression

We employ a linear model with ridge (L2) regression to predict rank outcomes of individual race results at a driver-constructor level. The input data comprises a sparse matrix of parent constructors and drivers. Ridge regression is chosen because it includes drivers with limited data by shrinking their coefficients, unlike L1 regularization which would remove them entirely. This allows us to maintain unique coefficients for each driver and constructor. This approach allows for unique coefficients representing each driver and constructor. The matrix equation for the ridge regression model is seen below in (2).

$$X\beta = Y$$

*X: Sparse matrix of drivers and constructors*

*β: Matrix of individual driver and constructor coefficients*

*Y: Rank finish for a given constructor-driver combination in a single race*

$$\begin{pmatrix} & \text{red\_bull} & \text{ferrari} & \text{mclaren} & \dots & \text{hamilton} \\ R_{2012, \text{ Austraila}} & 1 & 0 & 0 & \dots & 0 \\ R_{2012, \text{ Austraila}} & 0 & 1 & 0 & \dots & 0 \\ R_{2012, \text{ Austraila}} & 0 & 0 & 1 & \dots & 1 \\ \vdots & \vdots & \vdots & \vdots & \ddots & \dots \\ R_{2023, \text{ Abu Dhabu}} & 1 & 0 & 0 & \dots & 0 \end{pmatrix} \begin{pmatrix} \beta_{\text{red\_bull}} \\ \beta_{\text{ferrari}} \\ \beta_{\text{mclaren}} \\ \vdots \\ \beta_{\text{hamilton}} \end{pmatrix} = \begin{pmatrix} Y_{\text{red\_bull|driver}} \\ Y_{\text{ferrari|driver}} \\ Y_{\text{mclaren|hamilton}} \\ \vdots \\ Y_{\text{red\_bull|driver}} \end{pmatrix}$$

(2)

We iteratively run the regression for each race from 2012 to the present. For example, to predict the 2016 Monaco Grand Prix (2016 season, round 6), training data from all races from 2012 up to the 2016 Spanish Grand Prix (2016 season, round 5) are used. More recent results are weighted higher than older ones, as defined by the equations highlighted in (3), where constants are optimized against mean absolute error (MAE) – see Appendix A2 for hyperparameter selection detail. This exponential decay approach is borrowed from similar NBA player impact ratings, which utilize ridge regression (Mevedovsky).

$$\text{weight}_{\text{season}} = e^{0.75 * (\text{Current Season} - \text{Race Season})}$$

$$\text{weight}_{\text{round}} = e^{0.075 * (\text{Current Round} - \text{Race Round})}$$

$$\text{weight}_{\text{time decay}} = \text{weight}_{\text{season}} * \text{weight}_{\text{round}}$$

(3)

We combine the time-decay weights with the rank position weights described in section 2 as shown in (4).

$$\text{weight}_{\text{overall}} = \text{weight}_{\text{Rank Finish}} * \text{weight}_{\text{time decay}}$$

(4)

The use of a time-decayed ridge regression allows us to gain some of the benefits of a random effect model (Bell 2016), namely the shrinking of large low sample residuals and capturing point-in-time performance

### 3.2. Features

As discussed in the Introduction, the features used are a sparse matrix of parent constructors and drivers. Unlike Kesteren and Bell's work, which included weather conditions, we exclude weather due to its unpredictability before a race weekend. Circuit type (street versus permanent circuit) was considered but ultimately excluded for simplicity.

### 3.3. Coefficient Smoothing

As mentioned in section 3.1, we use the coefficients for a given constructor and driver as a measure of their impact over time. However, due to the nature of our time-decay weighting, the time-series view can be noisy, as seen in Figure 2 below.

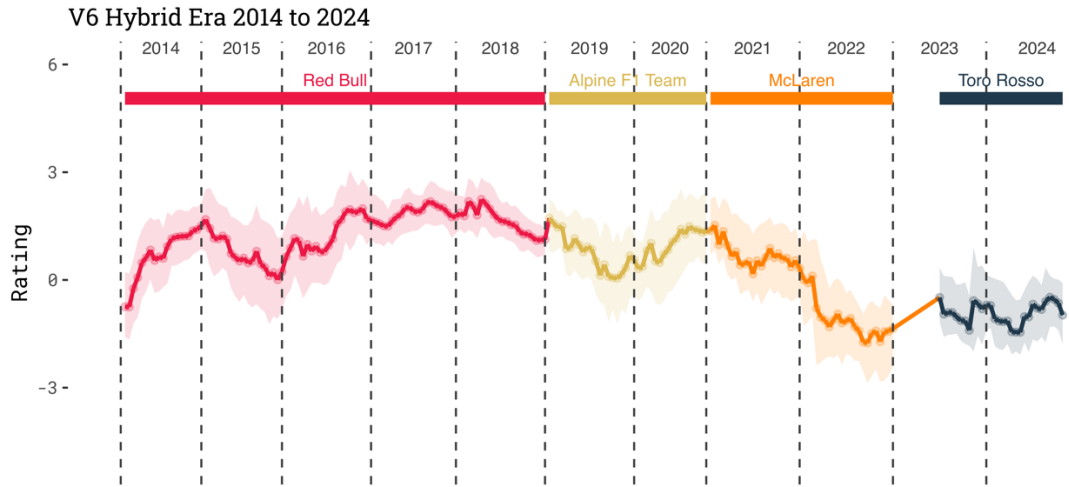

*Figure 2: Daniel Riccardo raw (model-derived) coefficient. The model-derived coefficient is particularly noisy at the start of Ricciardo's career, and each time he switches constructors. Shaded area represents a 95% confidence interval.*

To reduce this noise and improve predictive power, we use a blended average between the model-derived coefficients and a LOESS predicted coefficient for the next race. The blended average is defined by (5), with constants determined through hyperparameter optimization against mean absolute error (MAE).

$$\text{weight}_{\text{raw coefficient (rc)}} = \min(\frac{n_{\text{races}}}{40}, \begin{cases} 0.3, & \text{drivers} \\ 0.7, & \text{constructors} \end{cases})$$

$$\beta_{\text{Blended}} = \text{weight}_{\text{rc}} * \beta_{\text{raw}} + (1 - \text{weight}_{\text{rc}}) * \beta_{\text{LOESS}}$$

(5)

This smoothing methodology yields a less noisy and more accurate time series for constructor and driver rating as seen in Figure 3 below.

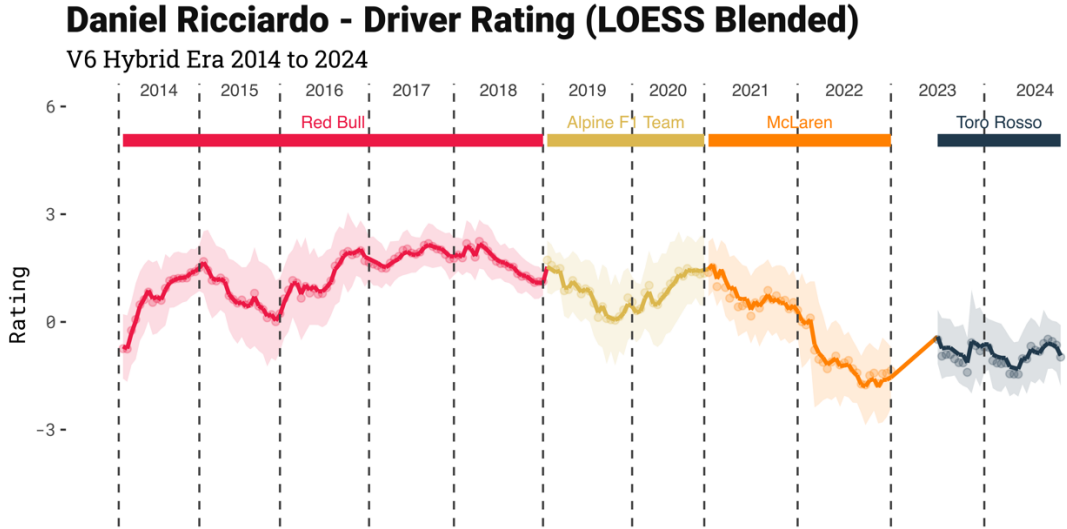

*Figure 3: Daniel Riccardo LOESS-blended coefficient. The LOESS blended coefficient is quick to pick up the initial performance surge at the start his career while maintaining stability in his driver coefficient later in his career. The transparent points represent the model-derived coefficient seen in Figure 2. Shaded area represents a 95% confidence interval.*

### 3.4. Bootstrapping

To measure confidence intervals for our coefficient measurements we apply a bootstrapping approach (R = 50) on each ridge regression. Assuming the individual coefficients are normally distributed within the bootstrapped replicates we arrive at a 95% confidence interval. That interval is represented by the shaded area in Figure 2 and 3.

However, it is important to note that we are solely measuring the variance of individual coefficients. That variance is likely understated for drivers and constructors with limited sample sizes as those coefficients will be penalized and shrunk through regularization. Additionally, penalized regression can result in non-normal distribution of those coefficients (Croiseau 2009). As such, the confidence intervals provided have limited value for sensitivity analysis but provide directional indicators to the variability of coefficients relatively to each other.

### 3.5. Testing

To understand the influence of constructors versus drivers, we decompose the variance explained by each component. We use Kendall's $\tau$ coefficient as our test metric due to its interpretability and greater robustness to outliers as compared to Spearman Rho. Robustness to outliers is particularly important in DNF-inclusive models. Alternatively, normalized discounted cumulative gain (nDCG) was considered due to its emphasis at the top of the ranking, however the inability to decompose driver and constructor impact would limit its interpretability.

#### 3.5.1. Test Data

For testing, we look at the 2014 to 2024 Formula 1 seasons. 2012 and 2013 seasons serve as a warm start for the model, while 2024 was the last completed Formula 1 season at the time of this analysis.

The test data is the immediate next race following the current race *r*. or for instance, the model from the 2016 Spanish Grand Prix (*model*$_{2016, \text{Round } 5}$) predicts the outcomes for the 2016 Monaco Grand Prix (2016 season, round 6). The predictions from the model are linearly ranked and used as $y_{pred}$. This same process is applied to get $y_{pred}$ for each model iteration *r* that we have. An out-of-sample test set that is immediately future-facing allows us to use a wholly predictive test method versus a backwards-facing out-of-sample test/train split.

### 3.5.2. Kendall's Coefficient

We calculate Kendall's coefficient at the race level and then average across all races in the test data as described in (6).

$$\tau_{\text{single race (r)}} = \frac{\text{Condordant Pairs (C)} - \text{Discordant Pairs (D)}}{\text{Combination}(n_{\text{drivers}}, 2)}$$

$$\tau_{\text{Overall}} = \frac{1}{R} \sum_{r=1}^{R} \tau_r$$

(6)

We extend Kendall's coefficient to calculate partial Kendall's Tau for constructors and drivers as described in (7).

$$\tau_{Partial,Constructors} = \tau_{y \mid Constructors \cdot Drivers} = \frac{\tau_{y \mid Constructors} - \tau_{y \mid Drivers} * \tau_{Drivers \mid Constructors}}{\sqrt{(1 - \tau_{y \mid Drivers}^2) * (1 - \tau_{Drivers \mid Constructors}^2)}}$$

$$\tau_{Partial,Drivers} = \tau_{y \mid Drivers \cdot Constructors} = \frac{\tau_{y \mid Drivers} - \tau_{y \mid Constructors} * \tau_{Drivers \mid Constructors}}{\sqrt{(1 - \tau_{y \mid Constructors}^2) * (1 - \tau_{Drivers \mid Constructors}^2)}}$$

$\tau_{y \mid Constructors}$: the Kendall's Tau between actual race positions (y) and constructor-only predictions

$\tau_{y \mid Drivers}$: the Kendall's Tau between actual race positions (y) and drivers-only predictions

$\tau_{Drivers \mid Constructors}$: the Kendall's Tau between constructors-only predictions and drivers-only predictions

(7)

From there, we can describe the proportional variance explained by drivers versus constructors.

$$Constructor\ Variance\ Explained = \frac{\tau_{Partial,Constuctors}^2}{\tau_{Partial,Constuctors}^2 + \tau_{Partial,Drivers}^2}$$

(8)

## 3.6. Logistic Model

In addition to the linear ridge regression, we designed a logistic model with ridge regression. This model uses the response variable of key positional benchmarks in Formula 1 corresponding to common sportsbook markets (Hennion 2023). The three binary response variables we use are top 3 (podium), top 6, and top 10 (points finish).

McFadden's R-squared is chosen as the test metric due to its interpretability, ease of implementation, and robustness.

$$R^2_{McFadden} = 1 - \frac{log(L_{model})}{log(L_{null})}$$

$$R^2_{McFadden,Partial,Constructors} = \frac{log(L_{drivers}) - log(L_{model})}{log(L_{null})}$$

$$R^2_{McFadden,Partial,Drivers} = \frac{log(L_{constructors}) - log(L_{model})}{log(L_{null})}$$

$log(L_{model})$ : Log-likelihood of the full model

$log(L_{constructors})$: Log-likelihood of the constructors only model

$log(L_{drivers})$: Log-likelihood of the drivers only model

$log(L_{null})$: Log-likelihood of the null model

(9)

We calculate the proportion of variance explained similarly to the base model with Kendall's coefficient as the test metric:

$$Constructor\ Variance\ Explained = \frac{R^2_{McFadden,Partial,Constructors}}{R^2_{McFadden,Partial,Constructors} + R^2_{McFadden,Partial,Drivers}}$$

(10)

# 4. Results

## 4.1. Linear Model Results

Table 1: Performance metrics for Time-Decay Ridge Regression Model. Kendall Tau is the primary test metric against predicted rank position of driver-constructors. Kendall's Tau is used over Spearman Rho due its robustness against outliers

| | **Time Weighted Linear Regression Model Performance** | | | | | | |
| | Tested on 2014 - 2024 Races | | | | | | |
| | **Model Metrics** | | | **Variance Explained** | | | |
| Model Data | MAE | Kendall Tau ($\tau$) | $\tau$-LOESS Blended | Partial $\tau_{Constructor}$ | Partial $\tau_{Driver}$ | Partial Constructor/Driver $\tau^2$ Ratio | Implied Constructor Influence |
| All DNFs Included | 3.4 | 0.529 | 0.524 | 0.305 | 0.171 | 3.179 | 76.1% |
| DNF Assigned to Constructor/Driver | 3.7 | 0.482 | 0.480 | 0.206 | 0.170 | 1.460 | 59.3% |
| No DNFs Included | 2.3 | 0.625 | 0.622 | 0.305 | 0.228 | 1.780 | 64.0% |
| Qualification | 3.0 | 0.595 | 0.593 | 0.286 | 0.304 | 0.886 | 47.0% |

Our model testing yields the metrics displayed in Table 1 above. Intuitively, the DNF-Excluded model outperforms the DNF-inclusive models, as removing DNFs largely removes outlier results. Notably, the LOESS-blended coefficients perform better than the raw coefficients.

The 64.0% variance explained by constructors in the DNF-excluded model contrasts with the 88% constructor-explained variance that Kesteren (2023) reported. While the intuitive thought is that the large difference is explained by the three additional years the proposed methodology covers, results show that is unlikely. Furthermore, the 2014-2021 timeframe perfectly encapsulates the 8 consecutive championships Mercedes won; 5 of which had Mercedes drivers finishing 1st and 2nd. However, when testing the ridge regression model on the same 2014-2021 cohort, we found an implied constructor influence of 70.1%. Consequently, we find only a small portion of the difference is explained through a difference in model data. The difference is largely explained through the difference in the model itself and the model testing, potentially due to the difference in model granularity – race vs season level.

In the DNF-inclusive model, 76.1% of the variance is explained by constructors, compared to 86% in Bell's DNF-inclusive model (Bell 2016). That difference can be likely explained through the wider time frame Bell uses (1979 - 2014). Classification percentage, a percentage of how many racers finished 90%+ of the race, has significantly increased over time, as shown in Figure 4 below. Consequently, constructor influence when looking at races before the V8 (2006 - 2013) era may be more of a measure of reliability than ability. We also observe a slight drop in classification percentage at the start of new engine regulations, suggesting the proposed model may lose predictive power with the new regulations set for the 2026 season.

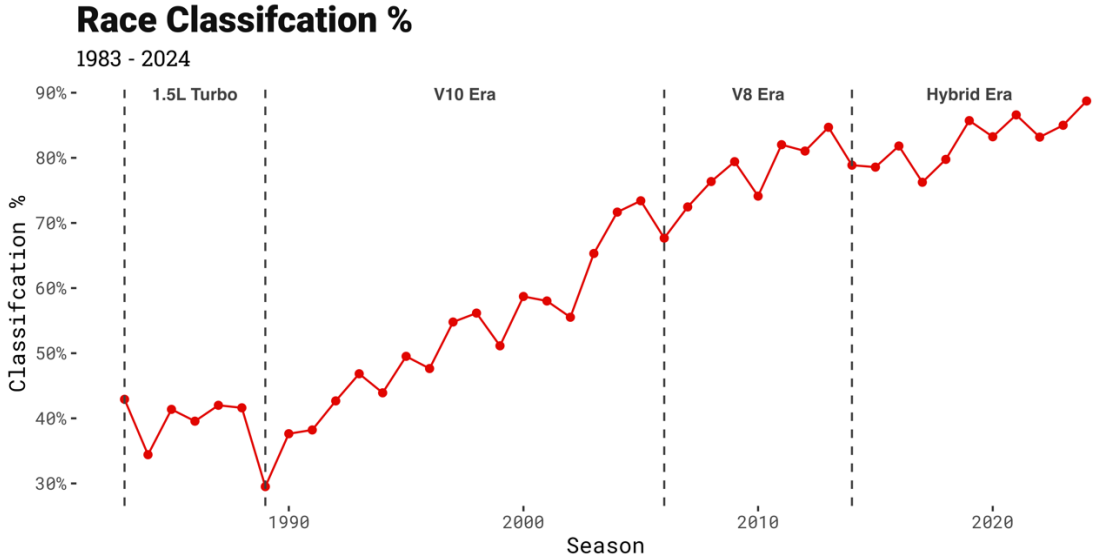

*Figure 4: Classification percentage by season, 1983 to present. Classified cars are those that finished 90% of a race. As measured through classification percentage, car reliability has steadily trended up over the last forty plus years. Consequently, DNF-inclusive approaches that use a larger time horizon may be a measure of reliability rather than ability*

Additionally, the decreased relative importance of constructors (47.0%) in predicting qualifying is noteworthy. The increased driver importance during qualification makes intuitive sense, as factors like race strategy and pit stops are essentially non-existent in qualifying compared to races.

### 4.2. Logistic Model Results

*Table 2: Performance metrics for Time-Decay Ridge Logistic Regression Model. McFadden's Pseudo R2 is the primary test metric. Constructor influence increases as N increases, however, model strength, as measured through McFadden's R2, decreases*

**Time Weighted Logistic Regression Model Performance (DNF-Inclusive)**
Against Top N Cut Offs | Tested on 2014 - 2024 Races

| Top N | Psuedo $R^{2[1]}$ | Partial Psuedo $R^2_{Constructor}$ | Partial Psuedo $R^2_{Driver}$ | Implied Constructor Influence |
|---|---|---|---|---|
| Podium (T3) | 0.294 | 0.088 | 0.025 | 78.3% |
| Top 6 | 0.204 | 0.084 | 0.024 | 77.7% |
| Points (T10) | 0.084 | 0.039 | 0.004 | 91.5% |

[1] McFaddens Psuedo $R^2$

Our logistic model against the 3 key sportsbook benchmarks (Top 3/6/10) suggests a larger constructor influence when examining broader cohorts compared to the rank-ordered predictions of the primary ridge regression model. The increased constructor importance suggests that constructors are the primary driver behind placing drivers into a given cohort (e.g. points, top 6), but the intra-cohort ordering is a larger function of driver ability.

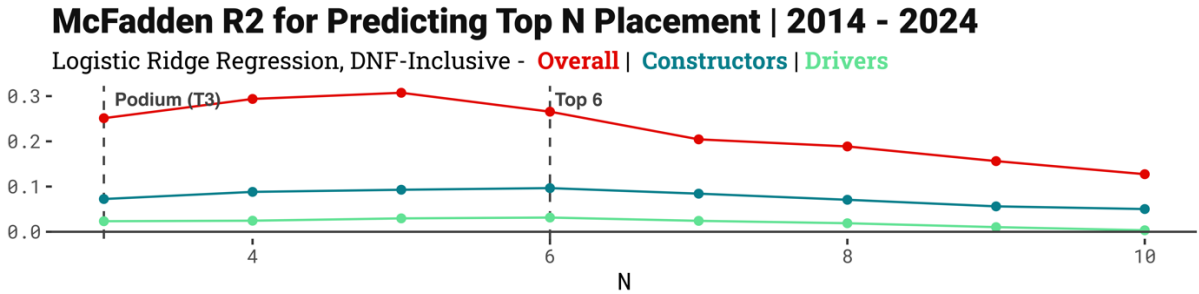

*Figure 5: Model Metrics for Top N Logistic Model. Model performance peaks at N = 5 before rapidly deteriorating for increasing N. The relative importance of constructors against drivers steadily increases for N greater than 6*

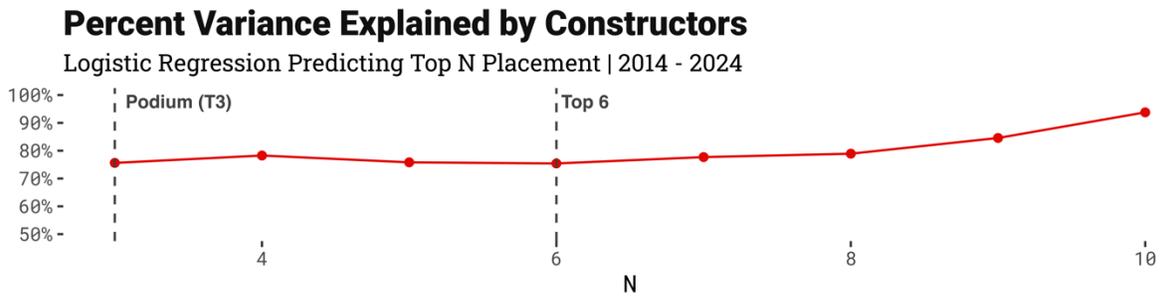

*Figure 6 Variance Explained for Top N Logistic Model. The relative importance of constructors against drivers steadily increases for N greater than 6*

We extend the logistic model for all Top N finishes for N 3 to 18. We find McFadden's $R^2$ dips below 0.0 for all N greater than 10. We see better model performance for lower values of N. Most interestingly, we see the importance of constructors in explaining variance steadily rise for increasing N, which supports the broader conclusion that constructors are increasingly important for placing within cohorts. Though, it is worth noting that model robustness declines for higher values of N. Additionally, we see a local maximum in model performance for N = 5 before model performance drops off at a higher rate. Notably we do not see a significant decline, if at all, in partial $R^2$ for both constructors and drivers for N < 7.

### 4.3. Driver and Constructor Coefficients

Analyzing driver and constructor coefficients provides additional context for intra-team driver performances and inter-team driver comparisons.

As shown below, we can better track Max Verstappen's model coefficient compared to his longtime teammate Daniel Ricciardo. Despite Ricciardo finishing with 92 points compared to Verstappen's 49 points in the 2015 season, their model coefficients were similar. The model attributed the point difference to constructor quality rather than driver quality. This trend continued over the next three seasons when the pair shared a team at Red Bull.

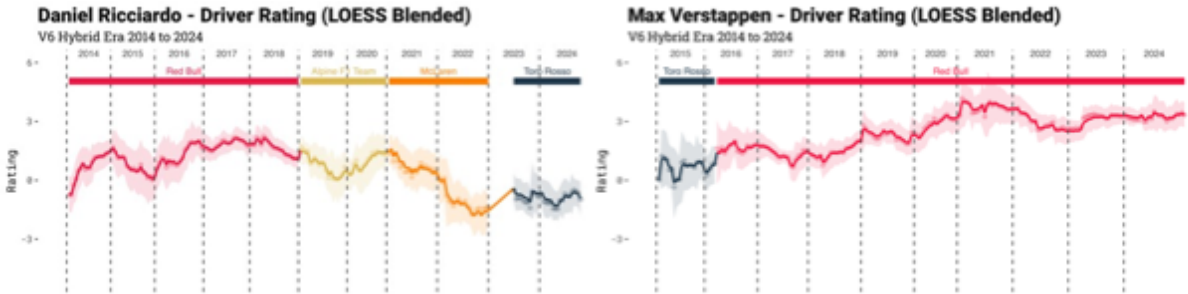

*Figure 7: Daniel Ricciardo vs Max Verstappen LOESS-Blended Coefficients/Ratings. Verstappen has a similar coefficient at the end of his 2015 Toro Rosso as Ricciardo's 2015 Red Bull Stint despite nearly doubling Verstappen's WDC total. The two drivers performed similarly in the 2016 season under the same constructor.*

A similar phenomenon is observed in comparing constructor coefficients. Mercedes is consistently seen as the dominant constructor throughout the 2014-2021 period, coinciding with their eight consecutive World Constructors' Championships.

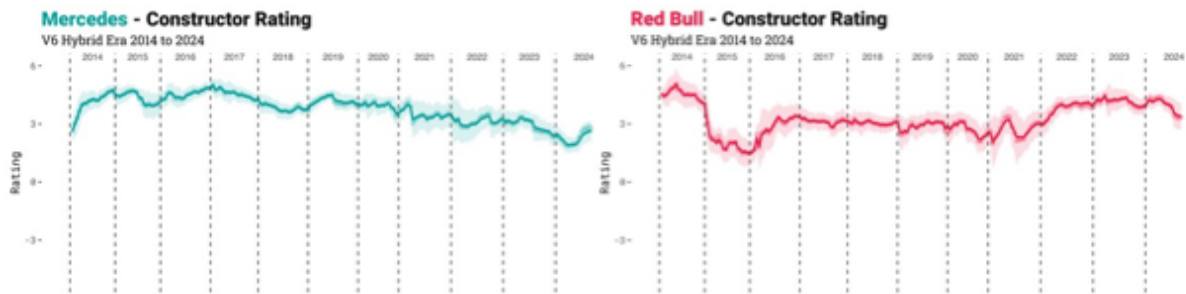

*Figure 8: Mercedes vs Red Bull LOESS-Blended Coefficients/Ratings. Mercedes has a higher coefficient until the 2022 season which is consistent with both teams' performances in the World Constructor Championships. Mercedes won the WCC from 2014 – 2021 while Red Bull won in 2022 and 2023. Shaded area represents a 95% confidence interval.*

While coefficients are useful for intra-team driver and constructor comparisons, similar conclusions can often be drawn from Formula 1 point totals. The back-and-forth between Verstappen and Ricciardo is evident in their point totals for the 2016-2018 seasons. Mercedes' dominance over Red Bull is also clear from their point totals relative to other constructors.

The unique insight provided by the ridge regression model lies in inter-team driver comparisons. This is illustrated with Max Verstappen, Lewis Hamilton, and Valtteri Bottas. Hamilton and Bottas finished 1st and 2nd, respectively, in both the 2019 and 2020 seasons, with Verstappen finishing 3rd in both seasons. Despite these results, Verstappen's driver coefficient for both seasons is higher than Bottas's and in line with Hamilton's.

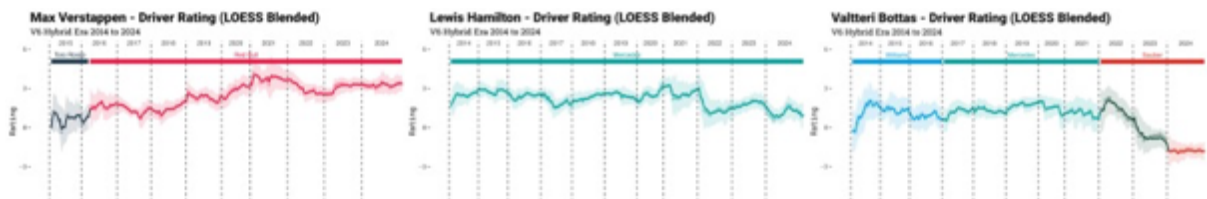

*Figure 9: Max Verstappen, Lewis Hamilton, Valtreri Bottas LOESS-Blended Coefficients/Rating. Verstappen has a higher coefficient than Bottas in the 2017 to 2020 seasons despite Bottas scoring more Formula 1 points in that time*

*frame, suggesting that the Bottas' performance was inflated by the strength of the Mercedes. Shaded area represents a 95% confidence interval.*

Bottas outperforming Verstappen in the World Driver's Championship is largely attributed to the quality of the Mercedes in the 2019 and 2020 seasons. Verstappen's talent was fully displayed in the 2021 season when he won the World Drivers' Championship as the gap between Mercedes and Red Bull closed. Hamilton and Bottas finished 2nd and 3rd respectively. While Bottas's rating is largely unchanged after leaving Wiliams for Mercedes, as expected, Bottas's steep drop off after leaving Mercedes for Sauber seems unlikely. The model may not be fully separating constructor and driver performance.

### 4.4. Alternate Methods

*Table 3 Model performance based on position weighting decision for DNF inclusive and exclusive models. No significant performance difference realized through current weighting system*

**Weighted/Unweighted Rank Model Performance**
Tested on 2014 - 2024 Races

| DNFs | Pos. Weighted | Kendall Tau ($\tau$) | Overall | T3 | T6 | T10 |
|---|---|---|---|---|---|---|
| Excluded | Pos. Weighted | 0.625 | 2.3 | 1.7 | 2.0 | 2.3 |
| Excluded | Pos. Unweighted | 0.630 | 2.3 | 1.7 | 2.0 | 2.3 |
| Included | Pos. Weighted | 0.482 | 3.7 | 3.2 | 3.5 | 3.6 |
| Included | Pos. Unweighted | 0.471 | 3.8 | 3.3 | 3.6 | 3.7 |

(Mean Absolute Error (MAE) covers the Overall, T3, T6, T10 columns)

While the design decision to weigh higher placed positions higher, the model performance remains largely independent of position weight. There may be future opportunities to further tune the weights to index heavier on points positions.

## 5. Discussion
### 5.1. Model Discussion

The overall DNF-Excluded model has a strong predictive power with a Kendall's coefficient of 0.625. This robustness instills confidence in the other derived conclusions. Tracking driver and constructor coefficients over time facilitates more informative inter-constructor comparisons.

The 64% variance explained by constructors, along with model metrics for qualification and logistic regression models, highlights the significant impact of constructors on driver finishing positions. The logistic model highlights the crucial role of constructors in determining cohort placements. Conversely, the qualification models emphasize the relative importance of drivers, suggesting potential value in incorporating qualification-informed models.

## 5.2. Model Extensibility

There are many opportunities to expand the modeling process outlined throughout this paper. For the sake of interpretability, the current model only includes drivers and (parent) constructors as features. Incorporating additional features, such as circuit type and separating the engine and constructor components, could enhance predictive accuracy. Moreover, the existing prediction process yields singular predictions. Implementing a bootstrapped model could provide probabilistic outcomes, thereby increasing the model's applicability to sports betting.

The presented methodology can also be extended to previous eras of Formula 1, allowing for cross-era comparisons. Expanding the model to cover earlier periods would likely necessitate recalibrating the time-decay factor to account for significant changes when engine regulations are altered. With new engine regulations set for the 2026 Formula 1 season, it will be essential to recalibrate the model to maintain its relevance and accuracy.

The methodology can be extended to other sports, particularly those related to racing. The methodology can largely be shifted over to horse racing, sailboat racing, and other motorsports. Extending the methodology to one-make (spec) racing where cars usually have the same chassis, powertrain, tires, brakes, and fuel would be an interesting contrast to the analysis presented in this paper. For example, we would expect almost all the variance to be explained by drivers in Formula 2 where all the chassis and engines are the same. Contrasting to semi-spec series such as IndyCar and Formula-E where chassis are the same, but power units can vary could provide a unique perspective on the impact of the driver.

## 5.3. Applications & Conclusions

*Table 4: Top 10 Constructor/Driver composite rankings as of the end of the 2023 F1 Season. Composite ranking is defined as driver rating plus constructor rating. ± values represent a 95% confidence interval. Driver and constructor independence is assumed for the overall confidence interval*

### Top 10 Driver/Constructor Composite Ratings
As of the Abu Dhabi Grand Prix, 2024

| Driver | Constructor | Driver Rating | Constructor Rating | Overall Rating | |
|---|---|---|---|---|---|
| Max Verstappen | Red Bull | +3.33 ± 0.32 | +2.46 ± 0.3 | +5.79 | ± 0.44 |
| Charles Leclerc | Ferrari | +1.92 ± 0.3 | +3.53 ± 0.27 | +5.45 | ± 0.4 |
| Lando Norris | McLaren | +2.00 ± 0.46 | +3.25 ± 0.34 | +5.25 | ± 0.57 |
| Carlos Sainz | Ferrari | +1.68 ± 0.24 | +3.53 ± 0.27 | +5.21 | ± 0.36 |
| Oscar Piastri | McLaren | +1.33 ± 0.4 | +3.25 ± 0.34 | +4.58 | ± 0.52 |
| George Russell | Mercedes | +1.72 ± 0.32 | +2.67 ± 0.28 | +4.39 | ± 0.43 |
| Lewis Hamilton | Mercedes | +0.96 ± 0.44 | +2.67 ± 0.28 | +3.63 | ± 0.52 |
| Sergio Pérez | Red Bull | −0.86 ± 0.46 | +2.46 ± 0.3 | +1.61 | ± 0.55 |
| Fernando Alonso | Aston Martin | +0.42 ± 0.33 | −1.10 ± 0.27 | −0.68 | ± 0.43 |
| Nico Hülkenberg | Haas F1 Team | +0.29 ± 0.5 | −1.23 ± 0.3 | −0.94 | ± 0.59 |

As mentioned in the Results section, the main applicability of the model is to discern driver impact amongst drivers on different teams. Additionally, we can capture point-time ratings for constructor/driver combinations to predict race outcomes as highlighted in the table above.

In conclusion, this study provides a nuanced understanding of the respective impacts of drivers and constructors on Formula 1 race outcomes. By leveraging a robust ridge regression model with LOESS smoothing and assessing predictive performance through Kendall's coefficient, we offer valuable insights into inter-constructor driver comparisons and the overall influence of constructors. The findings underscore the substantial role of constructors in determining race results, particularly in the context of modern Formula 1. The proposed methodology and findings pave the way for future research that can refine and extend these models, incorporating additional features and expanding to different eras in the sport. Such advancements hold promise for enhancing both predictive accuracy and practical applications, such as sports betting and strategic decision-making within Formula 1 teams.

## Acknowledgements

I would like to express my sincere gratitude to Akhil Raju and Naveen Venkatesan for their invaluable assistance as reviewers of this paper. Their insightful comments and constructive feedback greatly contributed to the improvement and clarity of the final manuscript.

## References

Bell, A., Smith, J., Sabel, C. and Jones, K. (2016) Formula for success: Multilevel modelling of Formula One Driver and Constructor performance, 1950–2014. Journal of Quantitative Analysis in Sports, Vol. 12 (Issue 2), pp. 99-112. https://doi.org/10.1515/jqas-2015-0050

Budzinski, O. and Feddersen, A. (2020) Measuring competitive balance in Formula One racing. In Outcome uncertainty in sporting events (pp. 5-26). Edward Elgar Publishing.

Croiseau, P. and Cordell, H.J., 2009, December. Analysis of North American Rheumatoid Arthritis Consortium data using a penalized logistic regression approach. In BMC proceedings (Vol. 3, pp. 1-5). BioMed Central.

Eichenberger, R. and Stadelmann, D. (2009) Who is the best Formula 1 driver? An economic approach to evaluating talent. Economic Analysis and Policy, 39(3), p.389.

Ergast. (n.d.). Ergast Developer API. Available at: https://ergast.com/mrd/methods/sprint/ (Accessed: 17 July 2024).

f1dataR. (n.d.). The f1dataR package. Available at: https://cran.r-project.org/web/packages/f1dataR/index.html (Accessed: 17 July 2024).

Formula 1. (n.d.). Formula 1. Available at: https://www.formula1.com/en.html (Accessed: 17 July 2024).

Fry, J., Brighton, T. and Fanzon, S. (2024) Faster identification of faster Formula 1 drivers via time-rank duality. Economics Letters, 237, p.111671.

Hennion, N. and Maher, T. (2023). How to Bet on Formula 1. Forbes. Available at: https://www.forbes.com/betting/formula-1/how-to-bet-on-formula-1/ (Accessed: 17 July 2024).

Jacobs, J. (2017). Deep Dive on Regularized Adjusted Plus-Minus I: Introductory Example. Squared2020. Available at: https://squared2020.com/2017/09/18/deep-dive-on-regularized-adjusted-plus-minus-i-introductory-example/ (Accessed: 17 July 2024).

Jacoby, W.G., 2000. Loess:: a nonparametric, graphical tool for depicting relationships between variables. Electoral studies, 19(4), pp.577-613.

Medvedovsky, K. and Patton, A. (2022) Daily adjusted and regressed Kalman optimized projections| DARKO. Accessed on, 17.

Moore, J., Dottle, R. and Paine, N. (2018). Formula One Racing. FiveThirtyEight. Available at: https://fivethirtyeight.com/features/formula-one-racing/ (Accessed: 17 July 2024).

Rockerbie, D. W. and Easton, S. T. (2022) 'Race to the podium: separating and conjoining the car and driver in F1 racing', Applied Economics, 54(54), pp. 6272–6285. doi: 10.1080/00036846.2022.2083068.

Sill, J., 2010, March. Improved NBA adjusted+/-using regularization and out-of-sample testing. In Proceedings of the 2010 MIT Sloan sports analytics conference.

Spurgeon, B. (2009) Age Old Question of Whether it is the Car or the Driver that Counts. Formula 1, 101.

van Kesteren, E. and Bergkamp, T. (2023) Bayesian analysis of Formula One race results: disentangling driver skill and constructor advantage. Journal of Quantitative Analysis in Sports, Vol. 19 (Issue 4), pp. 273-293. https://doi.org/10.1515/jqas-2022-0021

# Appendix

## A1. Companion Application & Data Download

A lightweight interactive React application is available at https://saurabhr.com/f1-rapm/ with ratings over time for all drivers and constructors from 2014 onwards and will continue to be updated with the latest data. Tabular data of the ratings can be download from the same portal.

## A2. Time Decay Selection

Table 5 Hyperparameter selection for time-decay season and round parameters. Season decay factor of 0.75 and round decay of 0.0750 was chosen.

**Time Weighted Linear Regression Model Performance**
On 2014 - 2024 Races

| Exponential Decay Factor[1] | | Mean Absolute Error | Kendall Tau ($\tau$) | Normalized Discounted Cumulative Gain | | |
|---|---|---|---|---|---|---|
| Season | Round | | | 3 | 6 | 10 |
| 0.7500 | 0.7500 | 2.49 | 0.60 | 0.69 | 0.79 | 0.82 |
| 0.7500 | 0.0750 | 2.29 | 0.63 | 0.73 | 0.82 | 0.85 |
| 0.7500 | 0.0075 | 2.37 | 0.61 | 0.72 | 0.81 | 0.84 |
| 0.0750 | 0.7500 | 2.50 | 0.60 | 0.69 | 0.79 | 0.82 |
| 0.0750 | 0.0750 | 2.33 | 0.62 | 0.73 | 0.82 | 0.84 |
| 0.0750 | 0.0075 | 2.55 | 0.58 | 0.68 | 0.78 | 0.81 |
| 0.0075 | 0.7500 | 2.49 | 0.60 | 0.69 | 0.79 | 0.82 |
| 0.0075 | 0.0750 | 2.34 | 0.62 | 0.73 | 0.82 | 0.84 |
| 0.0075 | 0.0075 | 2.59 | 0.58 | 0.68 | 0.78 | 0.81 |

[1] Overall decay weighting is:
$e^{(\text{Season Decay})*(\text{Current Season - Race Season})} * e^{(\text{Round Decay})*(\text{Current Round - Race Round})}$

The time-decay factor is chosen through a grid-search for both season and round level decay. A season decay factor of 0.75 and round decay of 0.075 was found to have the best performance by MAE. Kendall Tau and nDCG were used as additional guardrails for hyperparameter selection.

## A3. Metrics by Driver-Season Type

*Table 6: Mean Absolute Error (MAE) by Driver Type. Driver predictions in seasons after they have switched teams are less robust than predictions when the team has stayed the same.*

**Mean Absolute Error by Driver-Season**
Tested on 2014 - 2024 Races

| Driver | Driver-Seasons | Avg Finishing Position | MAE |
| --- | --- | --- | --- |
| Same Team | 2,523 | 8.3 | 2.29 |
| New Team | 736 | 9.2 | 2.47 |
| New Driver | 411 | 13.2 | 1.98 |

When stratifying driver-seasons by whether the are a new driver, joining a new team, or staying with the same team. As expected, the model is stronger at predicting the performance of teams staying with the same constructor than those that have switched teams. Counter-intuitively to that, we see that the error amongst new drivers altogether is better than the other categories. The lower error amongst that cohort can potentially by explained by the nature of the constructors that new drivers are signed at. New drivers are often going to back-markers or mid-field teams that serve as feeder teams to contenders (e.g. Toro Rosso → Red Bull, Haas → Ferrari, Williams → Ferrari).

# A4. Parent Constructor Mapping

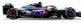

Figure 10: Mapping between parent constructors and constructors. Mapping allows coefficient consistency when team names change

# A5. Driver/Constructor Ratings as of 2024 British GP

Table 7: Driver/Constructor Composite Rankings as of the 2024 British GP. ± values represent a 95% confidence interval. Driver and constructor independence is assumed for the overall confidence interval

**Driver/Constructor Composite Ratings**
As of the Abu Dhabi Grand Prix, 2024

| Driver | Constructor | Driver Rating | Constructor Rating | Overall Rating | ± |
|---|---|---|---|---|---|
| Max Verstappen | Red Bull | +3.33 ± 0.32 | +2.46 ± 0.3 | +5.79 | ± 0.44 |
| Charles Leclerc | Ferrari | +1.92 ± 0.3 | +3.53 ± 0.27 | +5.45 | ± 0.4 |
| Lando Norris | McLaren | +2.00 ± 0.46 | +3.25 ± 0.34 | +5.25 | ± 0.57 |
| Carlos Sainz | Ferrari | +1.68 ± 0.24 | +3.53 ± 0.27 | +5.21 | ± 0.36 |
| Oscar Piastri | McLaren | +1.33 ± 0.4 | +3.25 ± 0.34 | +4.58 | ± 0.52 |
| George Russell | Mercedes | +1.72 ± 0.32 | +2.67 ± 0.28 | +4.39 | ± 0.43 |
| Lewis Hamilton | Mercedes | +0.96 ± 0.44 | +2.67 ± 0.28 | +3.63 | ± 0.52 |
| Sergio Pérez | Red Bull | −0.86 ± 0.46 | +2.46 ± 0.3 | +1.61 | ± 0.55 |
| Fernando Alonso | Aston Martin | +0.42 ± 0.33 | −1.10 ± 0.27 | −0.68 | ± 0.43 |
| Nico Hülkenberg | Haas F1 Team | +0.29 ± 0.5 | −1.23 ± 0.3 | −0.94 | ± 0.59 |
| Pierre Gasly | Alpine F1 Team | +0.44 ± 0.7 | −1.66 ± 0.46 | −1.22 | ± 0.84 |
| Franco Colapinto | Williams | +0.54 ± 0.46 | −2.54 ± 0.36 | −2.00 | ± 0.58 |
| Yuki Tsunoda | Toro Rosso | −0.23 ± 0.5 | −1.83 ± 0.34 | −2.07 | ± 0.61 |
| Jack Doohan | Alpine F1 Team | −0.80 ± 0.66 | −1.66 ± 0.46 | −2.46 | ± 0.81 |
| Liam Lawson | Toro Rosso | −0.67 ± 0.73 | −1.83 ± 0.34 | −2.51 | ± 0.8 |
| Kevin Magnussen | Haas F1 Team | −1.36 ± 0.49 | −1.23 ± 0.3 | −2.59 | ± 0.58 |
| Lance Stroll | Aston Martin | −1.50 ± 0.39 | −1.10 ± 0.27 | −2.60 | ± 0.47 |
| Alexander Albon | Williams | −0.28 ± 0.54 | −2.54 ± 0.36 | −2.82 | ± 0.65 |
| Guanyu Zhou | Sauber | −1.65 ± 0.32 | −3.50 ± 0.25 | −5.14 | ± 0.41 |
| Valtteri Bottas | Sauber | −1.83 ± 0.32 | −3.50 ± 0.25 | −5.33 | ± 0.41 |

*Table 8: Driver Rankings as of the 2024 British GP. ± values represent a 95% confidence interval*

## Driver Ratings
As of the Abu Dhabi Grand Prix, 2024

| Driver | Constructor | Driver Rating | Driver | Constructor | Driver Rating |
|---|---|---|---|---|---|
| Max Verstappen | Red Bull | +3.33 ± 0.32 | Nico Hülkenberg | Haas F1 Team | +0.29 ± 0.5 |
| Lando Norris | McLaren | +2.00 ± 0.46 | Yuki Tsunoda | Toro Rosso | −0.23 ± 0.5 |
| Charles Leclerc | Ferrari | +1.92 ± 0.3 | Alexander Albon | Williams | −0.28 ± 0.54 |
| George Russell | Mercedes | +1.72 ± 0.32 | Liam Lawson | Toro Rosso | −0.67 ± 0.73 |
| Carlos Sainz | Ferrari | +1.68 ± 0.24 | Jack Doohan | Alpine F1 Team | −0.80 ± 0.66 |
| Oscar Piastri | McLaren | +1.33 ± 0.4 | Sergio Pérez | Red Bull | −0.86 ± 0.46 |
| Lewis Hamilton | Mercedes | +0.96 ± 0.44 | Kevin Magnussen | Haas F1 Team | −1.36 ± 0.49 |
| Franco Colapinto | Williams | +0.54 ± 0.46 | Lance Stroll | Aston Martin | −1.50 ± 0.39 |
| Pierre Gasly | Alpine F1 Team | +0.44 ± 0.7 | Guanyu Zhou | Sauber | −1.65 ± 0.32 |
| Fernando Alonso | Aston Martin | +0.42 ± 0.33 | Valtteri Bottas | Sauber | −1.83 ± 0.32 |

*Table 9: Constructor Rankings as of the 2024 British GP. ± values represent a 95% confidence interval.*

**Constructor Ratings**

As of the Abu Dhabi Grand Prix, 2024

| Constructor | Constructor Rating | Constructor | Constructor Rating |
|---|---|---|---|
| Ferrari | +3.5 ± 0.27 | Haas F1 Team | −1.2 ± 0.3 |
| McLaren | +3.2 ± 0.34 | Alpine F1 Team | −1.7 ± 0.46 |
| Mercedes | +2.7 ± 0.28 | Toro Rosso | −1.8 ± 0.34 |
| Red Bull | +2.5 ± 0.3 | Williams | −2.5 ± 0.36 |
| Aston Martin | −1.1 ± 0.27 | Sauber | −3.5 ± 0.25 |

# A6. Driver/Constructor Coefficient/Ratings Trends 2014 - 2024 British GP

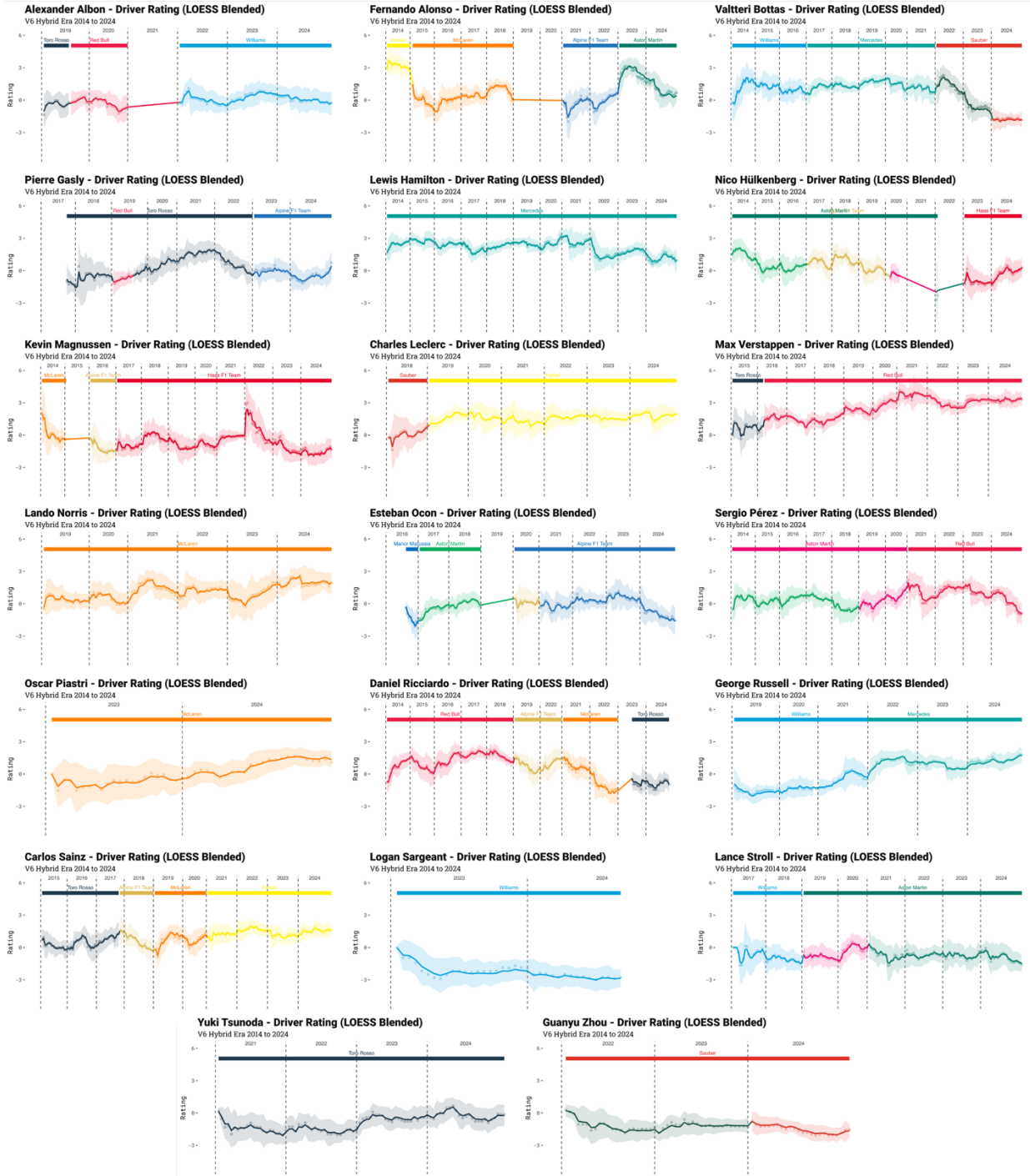

Figure 11: Driver Ranking/Coefficient Trend 2014 - 2024 British GP. Shaded area represents a 95% confidence interval.

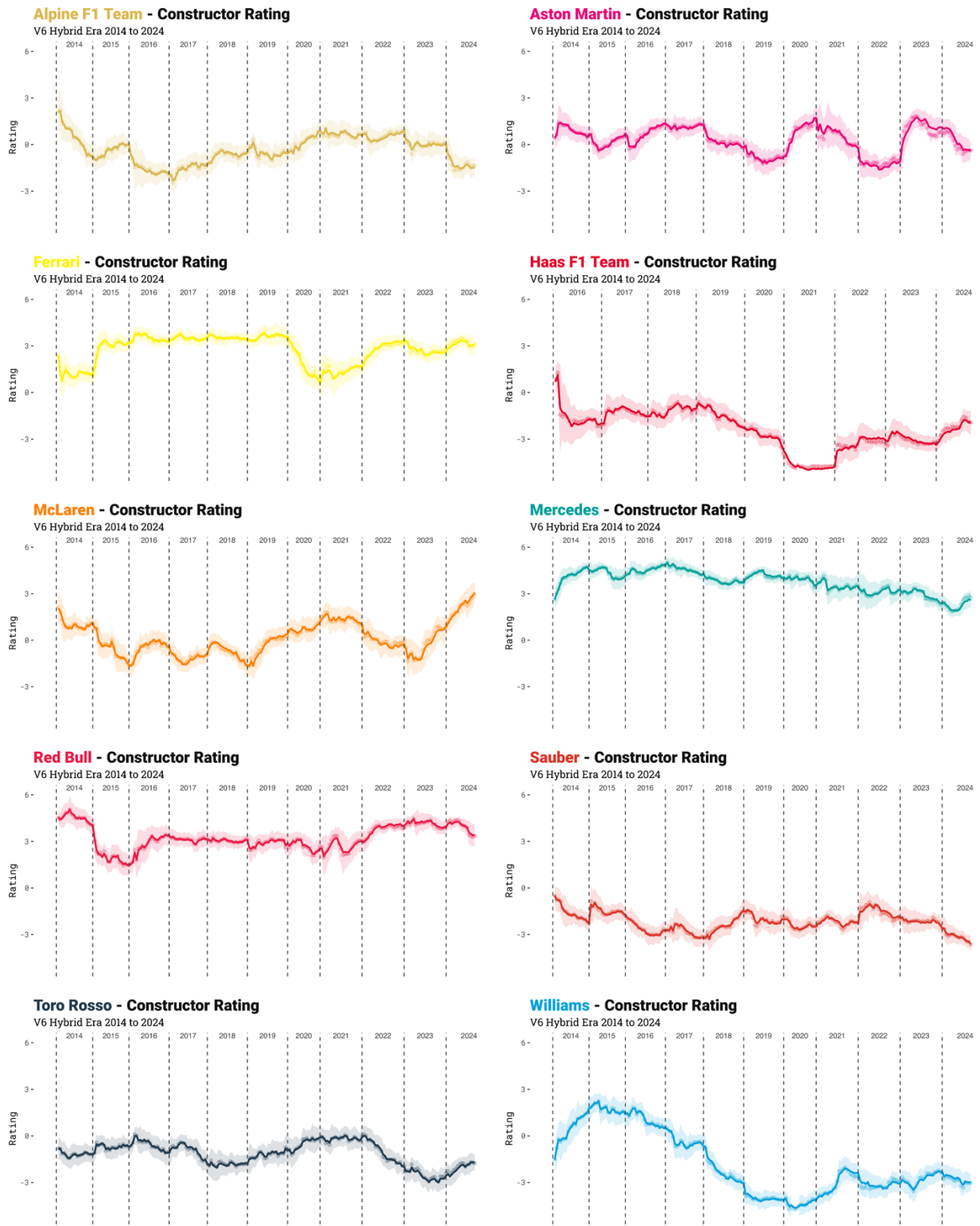

*Figure 12: Constructor Ranking/Coefficient Trend 2014 - 2024 British GP. Shaded area represents a 95% confidence interval.*